# Site Selectivity on Chalcogen Atoms in Superconducting La(O,F)BiSSe

Masashi TANAKA[1,a], Takuma YAMAKI[1,2], Yoshitaka MATSUSHITA[1], Masaya FUJIOKA[1],

Saleem J. DENHOLME[1], Takahide YAMAGUCHI[1], Hiroyuki TAKEYA[1],

and Yoshihiko TAKANO[1,2]

[1]National Institute for Materials Science, 1-2-1 Sengen, Tsukuba, Ibaraki 305-0047, Japan

[2]University of Tsukuba, 1-1-1 Tennodai, Tsukuba, Ibaraki 305-8577, Japan

[a]Corresponding author: Masashi Tanaka

E-mail: Tanaka.Masashi@nims.go.jp

Postal address: National Institute for Materials Science

1-2-1 Sengen, Tsukuba, Ibaraki 305-0047, Japan

Tel.: (+81)29-851-3354 ext. 2976

**Abstract**

Single crystals of La(O,F)BiSSe were grown by using a CsCl flux method. Single crystal X-ray structural analysis reveals that the crystal structure is isostructural to the $BiS_2$- or $BiSe_2$-based compounds crystallizing with space group *P4/nmm* (lattice parameters *a* = 4.1110(2) Å, *c* = 13.6010(7) Å). However, the S atoms are selectively occupied at the apical site of the Bi-SSe pyramids in the superconducting layer. The single crystals show a superconducting transition at around 4.2 K in the magnetic susceptibility and resistivity measurements. The superconducting anisotropic parameter is determined to be 34-35 from its upper critical magnetic field. The anisotropy is in the same range with that of other members of the La(O,F)Bi$Ch_2$ (*Ch* = S, Se) family under ambient pressure.

After the discovery of superconductivity in $Bi_4O_4S_3$,[1] various kinds of $BiS_2$-based superconductors have been developed, such as the $Ln$(O,F)$BiS_2$ ($Ln$ = La, Ce, Pr, Nd, Yb),[2-7] (La,$M$)O$BiS_2$ ($M$ = Th, Hf, Zr, Ti),[8] (Sr,La)F$BiS_2$ families.[9] They crystallize with a layered structure which is composed of superconducting $BiS_2$ layers and charge reservoir blocking layers similar to which is found for the cuprate or Fe-based superconductors.

La(O,F)$BiS_2$ is a typical $BiS_2$-based superconductor and comprises of the blocking layer La(O,F) as shown in Figure 1. The same blocking layer is also present for the Fe-based superconductor La(O,F)FeAs.[10] Recently, the modification of the superconducting layer has been explored in an attempt to design a new type of superconductor. The superconducting layer can be modified by changing the S atoms to Se. Polycrystalline samples of La(O,F)BiSSe ($T_c$ ~3.8 K)[11] and La(O,F)$BiSe_2$ ($T_c$ ~2.6 K)[12] have been reported. Among the Se-substituted system of La($O_{0.5}F_{0.5}$)Bi(S,Se)$_2$, the materials start to exhibit metallic behavior and $T_c$ shows its highest value at a ratio of S:Se = 1:1.[13] It is interesting to know why $T_c$ shows the highest value at the stoichiometric ideal of La(O,F)BiSSe. Single crystals of the end members La(O,F)$BiS_2$ and La(O,F)$BiSe_2$ have been prepared and the crystal structures were carefully determined using single crystal X-ray structural analysis.[14, 15] It is important to compare the single crystal structure of La(O,F)BiSSe to that of the end members so as to help design a new compound with a higher $T_c$.

In this study, we have prepared single crystals of La(O,F)BiSSe by a CsCl flux method and carried out various characteristic measurements including single crystal X-ray diffraction. Here, we discuss the bonding nature and superconducting properties of this compound.

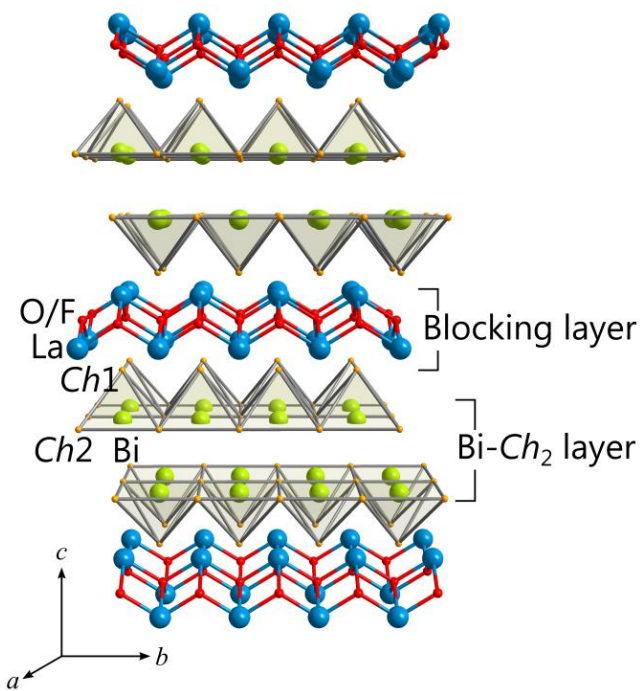

Fig. 1. Schematic illustration of the crystal structure of Bi$Ch_2$-based ($Ch$ = S, Se) superconductors.

Single crystals of La(O,F)BiSSe were prepared by an alkali metal chloride flux method similar to that of La(O,F)BiSe$_2$.[15] The synthesis details are described in the Supplementary Information.[16] Single crystal X-ray structural analysis was carried out using a Rigaku AFC11 Saturn CCD diffractometer with a VariMax confocal X-ray optical system for Mo K$\alpha$ radiation ($\lambda$ = 0.71073 Å). Prior to the diffraction experiment, the crystals were cooled to 113 K in a cold N$_2$ gas flow, in order to suppress the thermal displacement factors. Cell refinement and data reduction were carried out by the program d*trek package in CrystalClear software suite.[17] The structure was solved by the direct method using SHELXS,[18] and was refined with the program SHELXL[18] in the WinGX software package.[19] Energy-dispersive X-ray (EDX) spectra were observed using a scanning electron microscope (JEOL, JSM-6010LA). The temperature dependence of magnetic susceptibility was measured using a SQUID magnetometer (MPMS, Quantum Design) down to 1.8 K under a field of 10 Oe, and the field was applied parallel to the *c*-axis. The temperature dependence of electrical resistivity was measured down to 2.0 K, with a Physical Property Measurement System (PPMS, Quantum Design) using a standard four-probe method with constant current mode. The electrodes were attached in the *ab*-plane with silver paste.

Both the spot and overall EDX analysis measurements on the sample surface showed almost the same composition. The averaged compositional ratio of elements is estimated to be La : Bi : S : Se :

O : F = 1 : 0.95(8) : 0.92(13) : 0.9(2) : 0.32(4) : 0.19(2), in which the ratio was based on La = 1. The atomic ratio except for O and F elements is in good agreement with the nominal composition (LaO$_{0.5}$F$_{0.5}$BiSSe), suggesting that the single crystal has the composition La(O,F)BiSSe.

The obtained crystals are highly oriented along the *c*-axis (Fig.S1).[16] The X-ray single crystal refinement converged to the *R*1 values of 3.66 % and 4.37 % for $I \geq 2\sigma(I)$ and all data, respectively, for 16 variables including anisotropic parameters. The atomic coordinates and displacement parameters are given in TABLE I. The compound crystallizes with space group *P4/nmm* (lattice parameters *a* = 4.1102(2) Å, *c* = 13.6010(7) Å, and *Z* = 1). The crystal has an alternate stacking of BiSSe pyramids and La$_2$(O/F)$_2$ layers; the basic structure of this crystal is isostructural with La(O,F)BiS$_2$ and La(O,F)BiSe$_2$.[14,15] All atoms were located at special positions and their anisotropic displacement factors were all positive and within similar ranges, all of which appeared to be physically reasonable. Since the occupancy refinement in O and F under constrain of Occ.(O) + Occ.(F) = 1 did not converge in the refinement, the occupancies were fixed to the nominal composition, O:F = 0.5:0.5.

TABLE I. Atomic coordinates, atomic displacement parameters (Å$^2$), and bond valence sum (BVS) for the La(O,F)BiSSe.

| Atom | Wyck. | S.O.F | x/a | y/b | z/c | $U_{11}$[a] | $U_{22}$[a] | $U_{33}$[a] | $U_{eq}$[a] | BVS[b] |
|---|---|---|---|---|---|---|---|---|---|---|
| La | 2c | 1.0 | 3/4 | 3/4 | -0.09501(5) | 0.0079(2) | 0.0079(2) | 0.0111(3) | 0.0090(2) | +2.62 |
| Bi | 2c | 1.0 | 3/4 | 3/4 | 0.37196(3) | 0.00740(14) | 0.00740(14) | 0.0116(2) | 0.00879(12) | +3.36 |
| S/Se(Ch1) | 2c | S: 0.903(13) Se: 0.097(13) | 3/4 | 3/4 | 0.1858(2) | 0.0066(7) | 0.0066(7) | 0.0080(10) | 0.0070(6) | -1.90 / -0.16 |
| Se(Ch2) | 2c | 1.0 | 1/4 | 1/4 | 0.37860(11) | 0.0081(3) | 0.0081(3) | 0.0174(6) | 0.0112(2) | -2.21 |
| O/F | 2a | 0.5/0.5(Fix) | 3/4 | 1/4 | 0 | 0.007(2) | 0.007(2) | 0.010(3) | 0.0083(12) | -1.72 |

[a] $U_{12}$, $U_{13}$, and $U_{23}$ are 0, and $U_{eq}$ is defined as one third of the trace of the orthogonalized $U$ tensor.
[b] The BVS parameters $r_0$ and $B$ of are listed in supplemental material.[16]

Interestingly, the apical *Ch*1 site of the Bi-SSe pyramid is selectively occupied by S. It is reasonable to presume that a small amount of Se in *Ch*1 site is attributed to a migration of excess Se caused by the non-stoichiometry of the crystal. Namely, S and Se atoms are separately ordered in different sites in an ideal stoichiometric crystal as shown in Figure 2(a). This site selectivity suggests the possibility for designing a new superconducting layer by replacing the *Ch*1 and *Ch*2 atoms selectively.

Some selected bond distances and angles are given in Fig. 2(b) as the ORTEP representation and Table II in comparison with those of other structural analysis results on La(O,F)Bi*Ch*$_2$.[14,15] Six coordinated Bi atoms are bonded with S and/or Se atoms in three kinds such as Bi-*Ch*1, in-plane Bi-*Ch*2, and inter-plane Bi-*Ch*2 bondings. Focusing on the Bi-*Ch*1 bond, we can see a remarkable feature reflecting the site selectivity. As can be seen from TABLE II, for the two different samples with determined F-content, La(O$_{0.77}$F$_{0.23}$)BiS$_2$ and La(O$_{0.54}$F$_{0.46}$)BiS$_2$, respectively,[14] the Bi-*Ch*1 bonding length is almost the same value (with a 0.8 % difference in value). The distance also does not change so much when varying the chalcogen atoms from La(O,F)BiS$_2$ to La(O,F)BiSSe. However, it jumps up to more than 6 % when moving from La(O,F)BiSSe to La(O,F)BiSe$_2$. The difference of ~0.14 Å is identical to the difference between the ionic radii S$^{2-}$ (1.84 Å) and Se$^{2-}$ (1.98 Å).[20] This fact indicates that the in-plane *Ch*2 sites are selectively replaced by Se atoms rather than at the apical *Ch*1 sites. This tendency is reflected when the change in the *c*-lattice constant is plotted against increasing Se-content in the polycrystalline sample.[13] At a ratio of S:Se = 1:1 and beyond, there is a clear increase in the gradient of the slope. An observation that is not present in the case of *a*-axis.

Bond valence sum (BVS) was estimated from the observed bond distances and the BVS parameters ($r_0$ and *B* values) of nominal valence La$^{3+}$, Bi$^{3+}$, S$^{2-}$, Se$^{2-}$, O$^{2-}$, F$^-$ provided by Brown.[21] The BVS values are also listed in TABLE I. The BVS of Bi was estimated to be +3.36, indicating a difference of +0.36 from nominal valence. This result suggests the charge neutrality of this compound should be kept by modification of valence state of Bi as has been shown for the X-ray photoemission spectroscopy results for the BiS$_2$-based compound.[22]

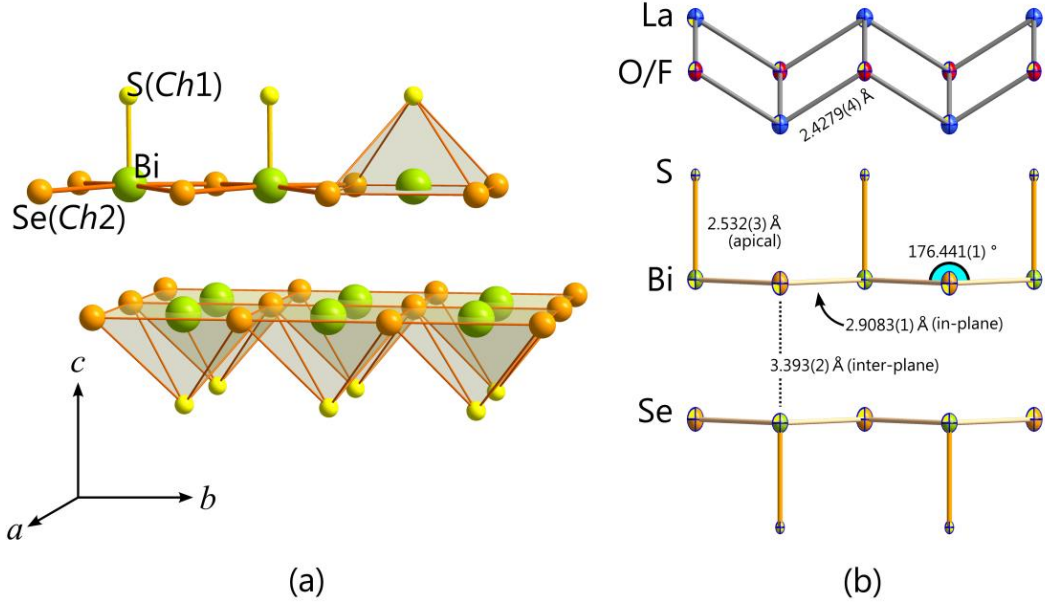

Fig. 2. (a) A schematic illustration of the local structure of the Bi-SSe pyramids in La(O,F)BiSSe. The chalcogen sites of *Ch*1 (apical) and *Ch*2 (in-plane) are selectively occupied by S and Se, respectively. (b) ORTEP representation of La(O,F)BiSSe with the selected bonding distances and angles. Displacement ellipsoids are drawn at an 80 % probability level.

TABLE II. Selected bond lengths (Å) and angles (°) of the La(O,F)BiSSe single crystal, compare to those of La(O,F)BiS$_2$ and La(O,F)BiSe$_2$.[14, 15] *Ch*1 and *Ch*2 denote the apical and in-plane sites on Bi*Ch*$_2$ pyramids, respectively.

| Material | La(O$_{0.77}$F$_{0.23}$)BiS$_2$ (Ref. 14) | La(O$_{0.54}$F$_{0.46}$)BiS$_2$ (Ref. 14) | La(O,F)BiSSe (This study) | La(O,F)BiSe$_2$ (Ref. 15) |
|---|---|---|---|---|
| Distance (Å) | | | | |
| La-O/F × 4 | 2.4117(16) | 2.4338(13) | 2.4279(4) | 2.4327(2) |
| La-*Ch*1 × 4 | 3.130(3) | 3.102(2) | 3.1584(10) | 3.2314(4) |
| Bi-*Ch*1 (apical) | 2.514(7) | 2.534(6) | 2.532(3) | 2.6752(8) |
| Bi-*Ch*2 (inter-plane) | 3.331(11) | 3.276(9) | 3.393(2) | 3.3245(9) |
| Bi-*Ch*2 (in-plane) × 4 | 2.870(2) | 2.8730(14) | 2.9083(1) | 2.9317(1) |
| Angle (°) | | | | |
| La-O/F-La | 114.51(9) | 113.17(8) | 115.687(1) | 116.660(8) |
| La-*Ch*1-La | 80.79(9) | 81.82(8) | 81.205(1) | 79.690(11) |
| *Ch*1-Bi-*Ch*2 | 91.38(19) | 89.59(16) | 91.779(8) | 92.869(14) |
| *Ch*2-Bi-*Ch*2 (in-plane) | 177.2(4) | 180.8(3) | 176.441(1) | 174.262(15) |

Figure 3 shows the temperature dependence of the magnetic susceptibility for a single crystal of La(O,F)BiSSe. The diamagnetic signal corresponds to superconductivity which appears below 4.1 K. The shielding and Meissner volume fraction at 1.8 K were estimated to be ~130 % and ~62 %, respectively. Since the magnetic field was applied parallel to *c*-axis in the plate-like shaped sample, the superconducting fraction includes overestimation due to its demagnetization effect.

Figure 4 shows the temperature dependence of the resistivity along the *ab*-plane in the single crystal. The resistivity shows metallic behavior above 7 K, and it deviates from the linear extrapolation of the normal part at around 5.0 K. Then it shows a sharp drop at around 4.2 K, and zero resistivity at around 3.7 K ($T_c^{zero}$), corresponding to the superconducting transition temperature $T_c$ found in the magnetic susceptibility measurement as shown in the inset. The onset temperature of superconductivity ($T_c^{onset}$) was determined to the temperature at 95 % of resistivity in its normal conduction state. The $T_c$ is clearly higher than that of the polycrystalline sample ($T_c \sim 3.8$ K). This tendency is also observed in La(O,F)BiSe$_2$.[15] It is likely that the high crystallinity is a factor in the observed $T_c$ for Bi-*Ch$_2$* based superconductors, although the reason why is unclear.

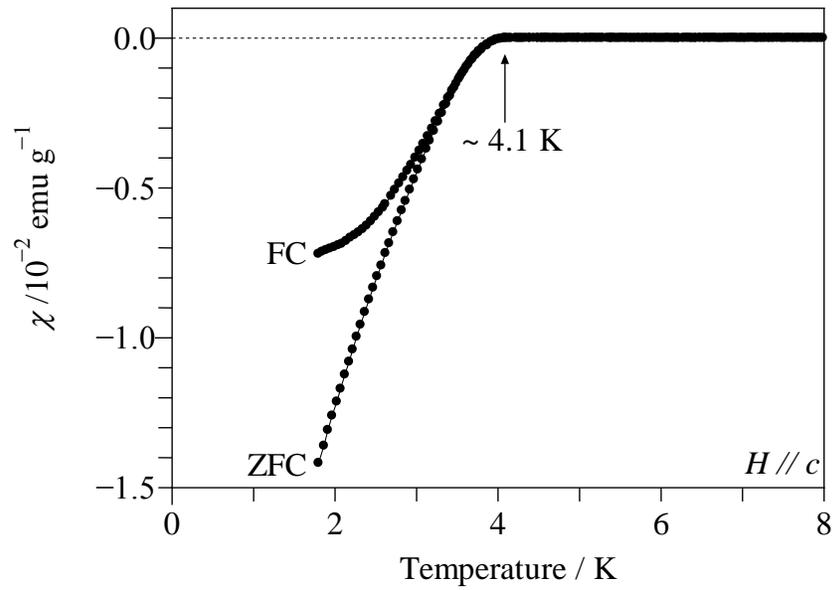

Fig. 3. Temperature dependence of magnetic susceptibility of the single crystal of La(O,F)BiSSe in zero-field cooling (ZFC) and field cooling (FC) mode. The magnetic field was applied parallel to *c*-axis.

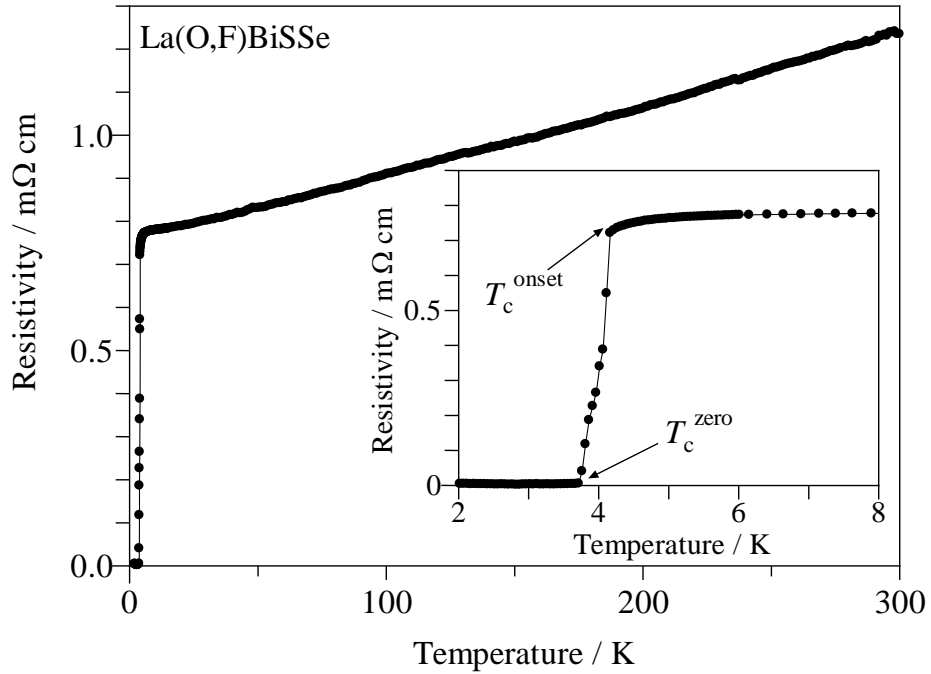

Fig. 4. Temperature dependence of resistivity of La(O,F)BiSSe. The inset shows an enlargement scale of the superconducting transition.

The resistivity was measured under various magnetic fields along the *ab*-plane (*H*//*ab*) and the *c*-axis (*H*//*c*) as shown in Figure 5(a) and (b), respectively. The superconductivity is significantly suppressed with increasing field parallel to *c*-axis in comparison to that in the case of *ab*-plane. Figure 6 shows the temperature dependence of the upper critical field ($H_{c2}(T)$), which was plotted from the values of $T_c^{onset}$ in Fig. 5. The $H_{c2}(0)$ in *H*//*ab* and *H*//*c* were estimated to be $H_{c2}^{//ab}(0) = 24.3$ and $H_{c2}^{//c}(0) = 0.7$ T, respectively, from the Werthamer-Helfand-Hohenberg (WHH) approximation for the Type II superconductor in a dirty limit. The linear extrapolation of $T_c^{onset}$ lead to $H_{c2}^{//ab}(0) = 37.4$ T and $H_{c2}^{//c}(0) = 1.1$ T. From these results, the superconducting anisotropic parameter $\gamma = H_{c2}^{//ab}(0) / H_{c2}^{//c}(0)$ is determined to be 34-35. The $\gamma$ of La(O,F)BiS$_2$ varies from 35 to 45 depending on its fluorine content.[23] In case of La(O,F)BiSe$_2$, the $\gamma$ value was reported to 20-49.[24, 25] $\gamma$ = 34-35 determined in this study is almost the same value with those of the other La(O,F)Bi*Ch*$_2$ superconductors.

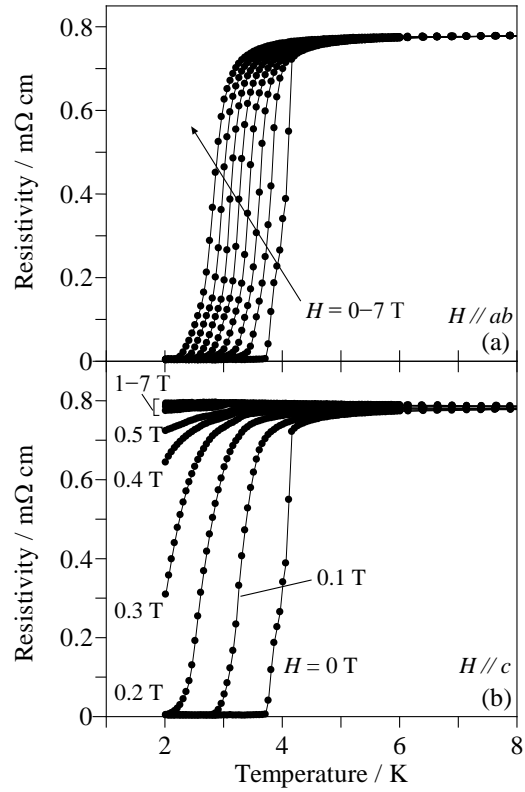

Fig. 5. Temperature dependence of resistivity of La(O,F)BiSSe under various magnetic field in (a) $H//ab$ and (b) $H//c$.

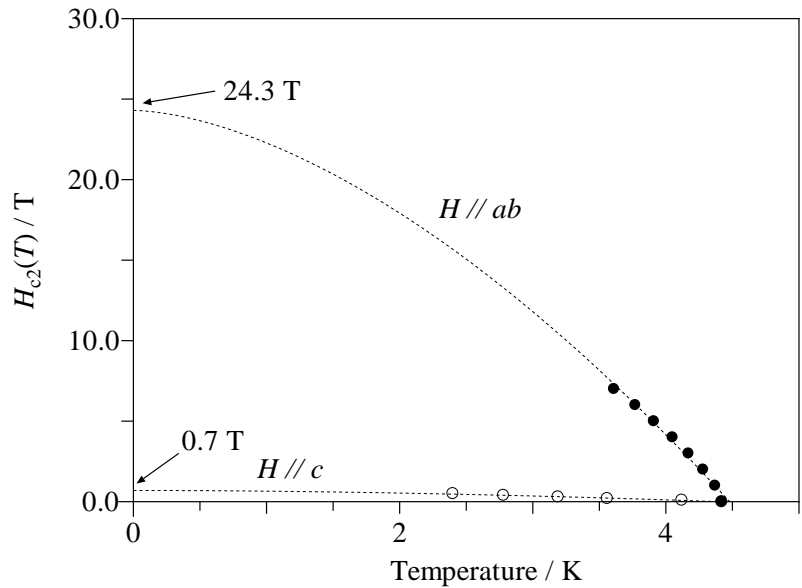

Fig. 6. Temperature dependence of upper critical field $H_{c2}(T)$ of the single crystal of La(O,F)BiSSe. The dotted line shows WHH approximation to estimate $H_{c2}(0)$.

Note that the $T_c$ is clearly higher than that of La(O,F)BiS$_2$ ($T_c$ = 2.6 K)[2, 14] and La(O,F)BiSe$_2$ ($T_c$ = 3.6 K).[12, 15] The relatively higher $T_c$ is attributed to the arrangement of apical S and in-plane Se for the Bi-SSe pyramids, which may enhance the density of states at the Fermi level. Theoretical studies are required to know the reason for the enhancement in $T_c$. And further to this, there is a possibility to achieve a higher $T_c$ by applying high pressure to La(O,F)BiSSe in a fashion analogous to that which has already been applied to other BiSe$_2$- or BiS$_2$-based materials.[25-27] This also should be investigated in future works.

In conclusion, single crystals of La(O,F)BiSSe have been prepared using the CsCl flux method. La(O,F)BiSSe crystalized with space group *P4/nmm* (lattice parameters $a$ = 4.1110(2) Å, $c$ = 13.6010(7) Å). The single crystal X-ray structural analysis revealed that the apical site of the Bi-SSe pyramid in the superconducting layer is selectively occupied by S atoms. The magnetic susceptibility and resistivity measurements showed a clear superconducting transition at around 4.2 K. The superconducting anisotropic parameter was determined to be 34-35 from its upper critical magnetic field.

**Acknowledgment**

This work was partially supported by Japan Science and Technology Agency through Strategic International Collaborative Research Program (SICORP-EU-Japan) and Advanced Low Carbon Technology R&D Program (ALCA) of the Japan Science and Technology Agency.